\pdfoutput=1
\documentclass[12pt]{iopart}

\usepackage{amssymb,amsthm,amstext,amsopn}
\usepackage{graphicx}
\usepackage{algorithm,algorithmic}
\usepackage{color}
\usepackage{url}

\newcommand{\N}{\mathbb{N}}
\newcommand{\R}{\mathbb{R}}

\newcommand{\E}{\mathbb{E}}

\newcommand{\argmin}{\operatorname*{arg\,min}}

\newcommand{\diag}{\mathrm{diag}}

\newcommand{\eps}{\varepsilon}
\renewcommand{\phi}{\mathbf{\varphi}}

\newcommand{\CO}{\mathcal{O}}

\newcommand{\CN}{\mathcal{N}}

\newcommand{\bfd}{\mathbf{d}}

\newcommand{\bfP}{\mathbf{P}}

\newcommand{\bfn}{\mathbf{n}}

\newcommand{\bfc}{\mathbf{c}}

\newcommand{\bfM}{\mathbf{M}}

\newcommand{\bfw}{\mathbf{w}}

\newcommand{\bfZ}{\mathbf{Z}}

\newcommand{\bfv}{\mathbf{v}}

\newcommand{\bfQ}{\mathbf{Q}}

\newcommand{\bfzero}{\mathbf{0}}
\newcommand{\bfone}{\mathbf{1}}

\newcommand{\x}{x}														
\renewcommand{\xi}[1]{\x_{#1}}								

\newcommand{\hf}{\frac{1}{2}}

\newlength\iwidth
\newlength\iheight

\begin{document}

\title[Bayesian framework for strain identification]{A Bayesian framework for molecular strain identification from mixed diagnostic samples}
\author{Lauri Mustonen$^1$, Xiangxi Gao$^1$, Asteroide Santana$^2$, \\ Rebecca Mitchell$^{1,3}$, Ymir Vigfusson$^{1,4}$ and Lars Ruthotto$^1$}
\address{$^1$ Department of Mathematics and Computer Science, Emory University, 400~Dowman Drive, Atlanta, GA 30322}
\address{$^2$ School of Industrial and Systems Engineering, Georgia Institute of Technology, 755~Ferst Drive, Atlanta, GA 30318}
\address{$^3$ Nell Hodgson Woodruff School of Nursing, Emory University, 1520~Clifton Road, Atlanta, GA 30322}
\address{$^4$ School of Computer Science, Reykjavik University, Menntavegi 1, Iceland, 101}
\ead{\{lauri.mustonen,xiangxi.gao,rebecca.mans.mitchell,\allowbreak ymir.vigfusson,lruthotto\}@emory.edu, asteroide.santana@gatech.edu}
\vspace{10pt}
\begin{indented}
\item[]July 7, 2018
\end{indented}

\begin{abstract}
We provide a mathematical formulation and develop a computational framework for identifying multiple strains of microorganisms from mixed samples of DNA. Our method is applicable in public health domains where efficient identification of pathogens is paramount, \emph{e.g.}, for the monitoring of disease outbreaks. We formulate strain identification as an inverse problem that aims at simultaneously estimating a binary matrix (encoding presence or absence of mutations in each strain) and a real-valued vector (representing the mixture of strains) such that their product is approximately equal to the measured data vector. The problem at hand has a similar structure to blind deconvolution, except for the presence of binary constraints, which we enforce in our approach. Following a Bayesian approach, we derive a posterior density. We present two computational methods for solving the non-convex maximum a posteriori estimation problem. The first one is a local optimization method that is made efficient and scalable by decoupling the problem into smaller independent subproblems, whereas the second one yields a global minimizer by converting the problem into a convex mixed-integer quadratic programming problem. The decoupling approach also provides an efficient way to integrate over the posterior. This provides useful information about the ambiguity of the underdetermined problem and, thus, the uncertainty associated with numerical solutions. We evaluate the potential and limitations of our framework \emph{in silico} using synthetic and experimental data with available ground truths.

\bigskip


\end{abstract}


\section{Introduction}
\label{sec:introduction}

Many public health programs, such as epidemiological surveillance, rely crucially on taking and processing biological samples to gather information.
Diagnostic samples can often contain multiple genetic variants, so-called \emph{strains}, of the same microorganism (\emph{e.g.}, bacteria, viruses, or parasites) resulting from mutations and adaptation.
For instance, blood samples of malaria patients may exhibit multiple concurrent strains of malaria contracted from bites of several parasite-infected mosquitoes or even a single bite of a mosquito carrying multiple genetic variants of the parasite~\cite{Galinsky2015}.
Different strains of pathogens can exhibit different characteristics that directly impact human health, potentially affecting the severity of illness, contagiousness, and resistance to classes of drugs~\cite{tenover1997}.
Thus,  accurate identification of strains within diagnostic samples is critical.
Epidemiological applications of strain identification include control effort evaluation, such as malaria reduction programs where different strains respond to different interventions, and tracing of outbreaks for pathogens that are similar to benign microorganisms, such as commensal (harmless) bacteria, or for which hosts may carry multiple pathogenic strains simultaneously, such as \textit {E.~coli} bacteria. 

Unfortunately, strain identification of mixed diagnostic samples -- samples with multiple strains of a pathogen -- remains particularly vexing.
The state-of-the-art approaches, detailed in the following section, have shortcomings that limit their applicability to these public health challenges.
Culture-based approaches are resource and labor intensive to be deployed at scale; metagenomic approaches thus far lack sufficient discriminatory power within a species, and direct polymerase chain reaction (PCR) diagnoses do not provide a sufficient strain-level resolution for epidemiological outbreak investigations or national surveillance programs. A recent method of Zhu \emph{et al.}~bypasses these problems, but critically depends on perfect knowledge of possible strain types -- a dictionary -- that is unrealistic to assume or generate for epidemiological surveillance in the field~\cite{Zhu2018}. 

There are two parts to overcome the strain reconstruction challenge for mixed samples: defining an appropriate laboratory procedure for converting the sample into a data vector and designing algorithms for disambiguating the pathogen strains from that data vector.
On the laboratory side, our approach uses fast, affordable and widely available biological diagnostic tools, 
specifically a combination of DNA barcoding~\cite{daniels2008} and whole genome multilocus sequence typing (wgMLST)~\cite{eurosurveillance,maiden2013,quainoo2017}, to produce short-read amplifications of independent genomic targets in the mixed sample. 
For each mixed sample, the deep sequencing platform in the procedure produces a measurement vector of locus-by-locus frequency information, denoting the percentage of mutations at every target location, or single-nucleotide polymorphism (SNP) site, in the amplified DNA sequences from the sample. 

In this work, we tackle the algorithmic side of the strain reconstruction problem using this measurement vector as the only input.
Specifically, we derive a mathematical formulation and computational framework for identifying distinct strains of microorganisms as well as their proportions from mixed diagnostic samples. 
From the measurement data vector, we infer the identity and frequency of the strains with a Bayesian inverse problem approach. 
We derive an expression for the posterior distribution and present numerical methods for computing \emph{maximum a posteriori} (MAP) estimates and integrals over the posterior distribution (Section~\ref{sec:math}).
The problem is underdetermined, and thus we give particular emphasis to quantifying uncertainty in the reconstruction.
We evaluate the potential and limitations of our methods \emph{in silico} using numerical examples on both simulated and experimental data (malaria strains) with available ground truths (Section~\ref{sec:experiments}). 
Our results suggest that strain sequences can be reconstructed from mixed samples with moderate to high fidelity for a range of input vectors, with important caveats that are influenced, \emph{e.g.}, by the input measurement errors and the true strain frequencies.

\section{Background}
\label{sec:bg}

Our approach to strain reconstruction depends on interpreting output from a particular biological pipeline that is applied to a sample. 
Before we formalize our algorithms that depend on these outputs, we provide some context about their possible application and a brief description of and references for possible experimental pipelines.

Traditionally, biological samples were cultured in the laboratory to isolate a single
microorganism. The isolate was then cultured to obtain a single strain
sample of the microorganism, which could be analyzed~\cite{maiden2013}.  If a sample was
expected to have multiple strains, then multiple subcultures were processed
to obtain samples of each strain.  
However, not all microorganisms are amenable to artificial culture. 
Moreover, some subtypes often grow better
than others, resulting in an unequal representation of the true subtypes in
the sample~\cite{randall}. 
The number of subtypes
detected can also be highly influenced by the number of subsamples cultured~\cite{dopfer2008assessing}.
This relationship can be seen, for example, in samples with low concentrations of minor strains,
such as drug-resistant bacteria that respond poorer to the culturing process (lower fitness) than wild-type bacteria. 
It is also evident in mixed samples with a high diversity of strains, such as 
samples of the \textit{P.~falciparum} malaria parasite with five or more strains in a
single sample.

Recently developed PCR-based DNA amplification
techniques allow one to diagnose pathogens directly from original samples,
alleviating the need of isolating individual strains in
culture~\cite{JacobEtAl2011,SacchiEtAl2011,tenover1997,maiden2013,quainoo2017}.
These PCR-based techniques
are not only considerably faster than culture-based diagnosis, but also
less expensive.  Furthermore, Langley \emph{et al.}~have shown that
PCR-based, culture-independent, diagnostic tests can be more sensitive and
are easier to perform than
traditional, culture-based approaches~\cite{LangleyEtAl2015}.

Despite the benefits of the PCR techniques, direct PCR-based diagnosis lacks information on strains at the level necessary for some important epidemiological studies, such as outbreak investigations or national surveillance programs.
These applications require more detailed microorganism resolution~\cite{tenover1997, quainoo2017} than provided by PCR-based diagnoses, which commonly focus on the clinically relevant species- or serotype-level identification.

To gain more information about strains, metagenomic approaches that evaluate all the DNA in a sample and screen for microorganisms of interest are being used to distinguish multiple genetic variants.
These techniques, however, generally require ample depth of coverage (copies of genomic area of interest) across large sections of the genome, or long, linked reads to provide sufficient discrimination of strains for our target epidemiological applications~\cite{vollmers}.  

To uniquely identify the targeted microorganisms within samples, our approach is instead to use a combination of DNA barcoding~\cite{daniels2008} and wgMLST diagnostic tools \cite{daniels2008,eurosurveillance, maiden2013,quainoo2017} that rely on coverage of specific, often unlinked, genomic targets to increase discriminatory power relative to metagenomic screening. 
An observation recently published by Zhu~\emph{et al.}~\cite{Zhu2018}, which we made concurrently and independently in our project, is that the locus-by-locus frequency information provided by the deep sequencing programs of the pipeline can be used to distinguish between strains in mixed samples.  
We focus on the analysis of the data created by these targeted and unlinked short-read
diagnostic techniques for the remainder of this paper.

\section{Methods}
\label{sec:math}
Motivated by practical lab pipelines for generating measurement data from mixed diagnostic samples,
we now detail the algorithmic part of the strain reconstruction problem: translating a vector of mutation frequencies at each location into strain genomic sequences and the relative proportions of each strain within the sample. 
We begin by specifying the problem, notation, and assumptions before deriving a Bayesian formulation of the ensuing inverse problem.
Following recommended guidelines~\cite{KaipioSomersalo2006,CalvettiSomersalo2007}, we model all quantities in the forward model as random variables, design prior distributions to incorporate prior knowledge, and use Bayes' formula to obtain a posterior distribution.
We will explore the posterior in three ways.
\begin{enumerate}
\item We address MAP estimation as a non-convex mixed-integer optimization problem and
present an efficient local optimization method based on block coordinate descent to compute local modes of the posterior.
The algorithm exploits the separability of the objective function and provides both deterministic running time and scalability with respect to the number of unknowns.
\item
We reformulate the MAP estimation problem as a convex binary constrained problem and present a method based on existing algorithms for obtaining the largest posterior mode.
Specifically, the method computes a \emph{global} minimizer and, unlike the first method, yields a certificate of optimality. While we observed that the running time can vary significantly between inputs, the method enables one to validate and calibrate the often more efficient local optimization method.
\item
We develop a computationally tractable method to quantify the uncertainty in the reconstructed strains and their frequencies.
In particular, we propose an efficient integration technique for the posterior density
that leverages the separability in the structure of the posterior.
\end{enumerate}

\subsection{Problem specification}
\label{sub:formalism}

Assume at first that the number of strains, denoted by $n$, is known, and consider measurements at $m$ locations in a DNA sequence.
At each location and for each strain, a mutation (relative to a reference strain) is either present, which can be encoded as $1$, or absent, which corresponds to $0$. This binary vector is called a molecular barcode in the biological literature~\cite{daniels2008}.
Now let $\bfd \in \R^m$ denote the actual measurement data which, for this example, represents the percentage of mutations at $m$ defined SNP sites in the DNA sequence.
If $\bfw \in \R^n$ is a vector containing the relative frequency of each strain and $\bfM \in \{0,1\}^{m \times n}$ is a binary matrix encoding the presence and absence of mutations, the forward problem can be written as
\begin{equation}\label{eq:fwd}
    \bfd = \bfM \bfw + \bfn,
\end{equation}
where $\bfn \in \R^m$ represents the inevitable measurement noise.
The goal of the inverse problem is to infer both $\bfM$ and $\bfw$ from the measurement data $\bfd$.
In other words, we aim to identify the strains and their respective proportions in a given sample.

Even in the absence of noise, \eref{eq:fwd} corresponds to an underdetermined and ill-posed inverse problem: There are $mn+n$ unknowns but only $m$ knowns.
For example, the problem is invariant to permutations of the columns of $\bfM$ and rows in $\bfw$.
On the other hand, with noisy measurements, the equation $\bfM \bfw = \bfd$ does not, in general, have a solution that satisfies the prescribed binary constraint for $\bfM$.

\paragraph{Multiplicity of infection.}
In practical applications, the number of strains $n$ -- often referred to as the multiplicity of infection (MOI) -- is usually unknown and often difficult to estimate~\cite{Galinsky2015}.
We will discuss below how to include the estimation for $n$ in the inverse problem.

\paragraph{Related problems.}
The inverse problem corresponding to~\eref{eq:fwd} arises also in signal processing and wireless communications when one tries to reconstruct binary source signals from a linear mixture that is formed with unknown mixing weights \cite{diamantaras2000,diamantaras2006cluster}.
More generally, in \emph{blind source separation} the aim is to recover the original (not necessarily binary) signals and a mixing matrix when multiple linear combinations are observed \cite{belouchrani1997,talwar1994,behr2017}.
Having a matrix measurement also leads to a \emph{non-negative matrix factorization} problem \cite{lin2007}, which can be equipped with binary constraints as well~\cite{slawski2013}.
Notice that although the number of unknowns increases if $\bfw$ is replaced with a matrix, the number of knowns increases as well and the problem actually becomes less underdetermined compared to our case.
The inverse problem corresponding to~\eref{eq:fwd} bears also similarities with blind deconvolution~\cite{YouKaveh1996,ChanWong1998,ChanWong2000} in the sense that both the linear operator $\bfM$ (compare to blurring operator) and the vector (compare to image) are unknown.
A common property for all abovementioned problems is that they are \emph{bilinear}, which for~\eref{eq:fwd} means being linear in $\bfM$ for a fixed $\bfw$ and linear in $\bfw$ for a fixed $\bfM$.
On the other hand, linear inverse problems in which binary and continuous variables are present arise in, \emph{e.g.}, groundwater flow~\cite{LairdEtAl2006}.

If the noise $\bfn$ in~\eref{eq:fwd} is zero (or small enough) and if the matrix $\bfM$ contains sufficiently many distinct rows, then in most cases the vector $\bfw$ can be easily solved by sorting the values in the measurement $\bfd$ and assigning them to $\bfw$ while discarding values that are binary combinations of already assigned values \cite{diamantaras2000,diamantaras2006cluster,behr2017}.
However, in our case, there is no guarantee that the matrix contains enough distinct rows for this approach to work.
Our measurements also contain noise, which is why we take the Bayesian approach for the inverse problem.

\subsection{Generalization to multiple classes}
\label{sub:generalization}

We start by considering a generalization of \eref{eq:fwd} where each site in a strain can represent more than two classes 
of measured values (mutation vs.~no mutation in the binary case above), allowing 
modeling of SNPs that have multiple alternative options, insertions or deletions of genomic content,
or multiple linked differences within a target location, to name a few examples.
One immediate use of this generalization is to obtain more detailed strain reconstruction consisting of all four nucleotides found in DNA molecules:\ adenine, guanine, cytosine, and thymine.
Even with multiple classes, we formulate the generalized problem by using a binary matrix and real-valued vectors, and the computational methods will thus remain similar to the binary case described above.

As before, let $m$ denote the number of SNP sites (\emph{i.e.}, measurement locations) and $n$ the number of strains, and let $p \geq 2$ now denote the number of classes each strain can be associated with at each measurement location.
The measured data represents the frequencies of the classes at each location.
Because the frequencies sum up to one, the measurement can be represented using $q := m(p-1)$ elements.

The measurement $\bfd$ is interpreted as a block vector having $m$ blocks of size $(p-1)$.
Within each block, the first value corresponds to the frequency of the second class, the second value to that of the third class, and so on.
In this way, the pure binary case $p=2$ will be handled naturally, and in general, the frequency of the first class is just one minus the sum of the frequencies of other classes.
The matrix $\bfM$ defining the strains now becomes a binary matrix with $m$ blocks of size $(p-1) \times n$.
One can think of replacing the zeros and ones of the pure binary case with vectors $\bfzero \in \R^{p-1}$ and Euclidean $(p-1)$-dimensional unit vectors, respectively.
As an example, if $p=4$, then a strain is characterized at one measurement location by either $(0,0,0)^\top$, $(1,0,0)^\top$, $(0,1,0)^\top$, or $(0,0,1)^\top$.

\subsection{Deriving the posterior}
\label{sub:deriving_the_posterior}

The observations taken from the sample can be written as a bilinear forward problem
\begin{equation*}
    D = M W + N,
\end{equation*}
where the measurement, $D$, is a random variable of length $q$, $M$ is a random variable of size $q \times n$ with binary entries whose columns represent the different strains, and $W$ is a random variable of length $n$ with real entries between $0$ and $1$ that model the proportions in which different strains are present in the sample.
For each realization $\bfw$ of $W$ we therefore know that $\sum_{j=1}^n \bfw_j = 1$.

In this work, we assume that the additive noise, $N$, is a multivariate Gaussian random variable with vanishing mean $\bfzero \in \R^q$ and a known diagonal covariance matrix $\Gamma = \mathrm{diag}(\gamma_1^2, \gamma_2^2, \ldots, \gamma_q^2) \in \R^{q\times q}$. In other words, $\bfn$ in \eref{eq:fwd} is a realization of $N \sim \CN(\bfzero,\Gamma)$. This assumption is motivated by its simplicity but can also be justified when the data features a relatively high signal to noise ratio.
Several other noise models are possible -- note that data is obtained by a counting process -- and will be investigated in future work.
With Gaussian noise, the \emph{likelihood} of the observation, $\bfd$, given some fixed realizations $\bfM$ and $\bfw$ is
\begin{equation*}
     \pi(\bfd \mid \bfM, \bfw) = \left( \frac{1}{2\pi |\Gamma|}\right)^{q/2} \exp \mathopen{}\left( - \hf \| \bfM \bfw - \bfd \|^2_\Gamma \right)\mathclose{},
 \end{equation*}
where for a vector $\bfv \in \R^q$ we define  $\| \bfv\|^2_{\Gamma} := \bfv^\top \Gamma^{-1} \bfv$ and denote the determinant of the noise covariance by $|\Gamma| = \prod_{i=1}^q \gamma_i^2$.
Note that this simple noise model gives positive probability to negative observations, as well as observations where the sum of one block is greater than one, although in practice such observations should not exist.
In the following, we will drop the normalization constant from the probability densities for readability.

We use prior distributions to incorporate \emph{a priori} knowledge (\emph{i.e.}, knowledge uninformed by the data) on the distributions of $M$ and $W$.
In particular, the priors for $M$ and $W$ are assumed to be mutually independent so that the joint prior can be written as $\pi(\bfM,\bfw) = \pi(\bfM) \pi(\bfw)$.

The feasible set for the binary matrix $M$ is
\begin{equation*}
    \Omega_M := \left\{ \bfM \in \{0,1\}^{q \times n} : \ \bfM = (\bfM_1, \ldots, \bfM_m)^\top,\  \bfM_k \in \tilde{\Omega}_M, 1 \leq k \leq m \right\},
\end{equation*}
where each block belongs to the set
\begin{equation*}
    \tilde{\Omega}_M := \left\{ \bfM \in \{0,1\}^{(p-1) \times n} \ : \ \sum_{i=1}^{p-1} \bfM_{i,j} \leq 1, \ 1 \leq j \leq n \right\}.
\end{equation*}
In other words, the column sums of each block are at most $1$, since a strain cannot be associated with more than one class at each SNP site.
It is easy to see that the cardinality of $\tilde{\Omega}_M$ is $|\tilde{\Omega}_M| = p^n$, and thus $|\Omega_M| = p^{mn}$.
Henceforth, we will choose the prior distribution of $M$ to be uniform, \emph{i.e.}, $\pi(\bfM) \propto \chi_{\Omega_M}(\bfM)$, where $\chi_{\Omega_M}$ denotes the characteristic function of the set ${\Omega_M}$.
However, our numerical methods can be readily generalized to support different distributions.
For example, in Sections \ref{sub:map_estimation} and \ref{sub:integration} we could choose any distribution that is separable in the sense that
\begin{equation*}
    \pi(\bfM) \propto \chi_{\Omega_M}(\bfM) \exp\mathopen{}\left( -\hf\sum_{k=1}^m r_k(\bfM_k) \right)\mathclose{},
\end{equation*}
where $r_k$ is a function depending only on the $k$th block of $\bfM$.

To reduce the ambiguity arising from different permutations of the columns in $\bfM$, we assume in this work that the entries in the vector of proportions $W$ have non-increasing order. To be specific, we assume that $W$ is supported in the set
\begin{equation*}
    \Omega_W :=\left\{ \bfw \in \R^n \ : \ \sum_{j=1}^n \bfw_j = 1, \quad 1 \geq \bfw_1 \geq \bfw_2 \geq \ldots \geq \bfw_n \geq 0  \right\},
\end{equation*}
which is a subset of an $(n-1)$-dimensional affine hyperplane; see visualizations in Figures~\ref{fig:4by3} and \ref{fig:errortriangles} for the case $n=3$.
For simplicity, we assume that $W$ is uniformly distributed in $\Omega_W$, thus the prior density is $\pi(\bfw) \propto \chi_{\Omega_W}(\bfw)$.
Again, the setting can be readily generalized for other prior distributions.
In particular, assuming that $W$ is a truncated Gaussian random variable with mean $\overline{\bfw} \in \Omega_W$ and a positive-definite covariance matrix $\Gamma_W \in \R^{n \times n}$, \emph{i.e.},
\begin{equation*}
    \pi(\bfw) \propto \chi_{\Omega_W}(\bfw) \exp \mathopen{}\left( - \hf \| \bfw - \overline{\bfw}\|^2_{\Gamma_W} \right)\mathclose{},
\end{equation*}
would not add any difficulties in the algorithms that follow.

Having discussed both the likelihood and prior terms, we apply Bayes' formula
\begin{equation*}
    \pi(\bfM,\bfw \mid \bfd) = \frac{\pi(\bfd \mid \bfM,\bfw) \ \pi(\bfM,\bfw)}{\pi(\bfd)}
\end{equation*}
which for our choices for the priors leads to the posterior distribution
\begin{equation}\label{eq:posterior}
    \pi(\bfM,\bfw \mid \bfd) \propto \exp \mathopen{}\left( - \hf \| \bfM \bfw - \bfd \|^2_\Gamma\right)\mathclose{} \ \chi_{\Omega_M}(\bfM) \ \chi_{\Omega_W}(\bfw).
\end{equation}
The posterior probability encodes both information provided by the data and by our prior knowledge about the biological applications at hand.

\subsection{Block coordinate descent method for MAP estimation}
\label{sub:map_estimation}
Maximum a posteriori (MAP) estimation aims at finding the largest mode of the posterior distribution~\eref{eq:posterior}.
Taking the negative logarithm of the posterior density and denoting
\begin{equation*}\label{eq:varphi}
\phi(\bfM,\bfw) :=  \left\| \bfM \bfw - \bfd\right\|^2_{\Gamma}
\end{equation*}
we obtain the constrained minimization problem
\begin{equation}\label{eq:MAP}
    \min_{\bfM, \bfw} \ \phi(\bfM,\bfw) \quad \text{ subject to } \quad \bfM \in \Omega_M, \ \bfw \in \Omega_W.
\end{equation}
Problem~\eref{eq:MAP} is a \emph{mixed-integer nonlinear programming} (MINLP) problem due to the binary constraints on $\bfM$ and the bilinear term $\bfM \bfw$.

Solving MINLP problems is known to be challenging as the computational complexity grows, in general, exponentially with the number of binary variables~\cite{BelottiEtAl2013}.
We will use the bi-linearity of the forward problem and the separability of the posterior to obtain a block coordinate descent method whose computational cost is $\CO(m p^n)$, \emph{i.e.}, the complexity grows linearly with the number of measurement locations and exponentially with the number of strains, which for the strain disambiguation application is rather small due to biological implausibility of a host simultaneously harboring dozens of competing pathogen strains.
However, the method typically converges to a local minimum and may thus be needed to run several times to obtain a global minimizer; see also visualization of local minima in Figure~\ref{fig:4by3}.

The block coordinate descent method decouples the problem~\eref{eq:MAP} into two steps. The general idea is to alternate between updating the binary matrix $\bfM$ and the frequency vector $\bfw$ while keeping the respective other variable, or \emph{block}, fixed. This is equivalent to maximizing the probabilities $\pi(\bfM \mid \bfw,\bfd)$ and $\pi(\bfw \mid \bfM, \bfd)$ repeatedly.
At the $i$th iteration, starting from $(\bfM^{i-1}, \bfw^{i-1})$, we solve the two subproblems
\begin{eqnarray}
    \bfM^{i} &= \argmin_{\bfM} \phi(\bfM, \bfw^{i-1}) \quad  & \text{ subject to }\quad \bfM \in \Omega_M \label{eq:getM},\\
    \bfw^{i} &= \argmin_{\bfw} \phi(\bfM^{i}, \bfw) \quad &\text{ subject to }\quad \bfw \in \Omega_W. \label{eq:getW}
\end{eqnarray}
For $p=2$, this technique is presented in \cite{talwar1994,talwar1996,slawski2013}, and a similar alternating minimization approach has also been successfully employed in blind deconvolution~\cite{YouKaveh1996}.

In the first step, we find an exact solution to the binary-constrained optimization problem~\eref{eq:getM}. 
Na\"ive solution of this problem would require full enumeration of all $p^{mn}$ possible matrices and would be prohibitively expensive.
However, we can decouple the problem along the blocks of the matrix, which yields $m$ independent problems
    \begin{equation}\label{eq:getMk}
        \bfM_{k}^{i} = \argmin_{\bfM_{k}} \left\| \bfM_{k} \bfw^{i-1} - \bfd_k\right\|^2_{\Gamma_k} \quad \text{ subject to } \quad  \bfM_k \in \tilde{\Omega}_M, 
    \end{equation}
where $\bfd_k \in \R^{p-1}$ denotes the $k$th block of the measurement vector for $ k=1,2,\ldots,m$, and $\Gamma_k \in \R^{(p-1) \times (p-1)}$ is the corresponding block in the noise covariance matrix.
Solving~\eref{eq:getMk} can be parallelized, giving rise to additional computational savings.
For the small problem sizes arising in the motivating public health applications, we use a full enumeration to solve each subproblem. Nevertheless, efficient software libraries such as Minotaur~\cite{leyffer2012minotaur} and Gurobi~\cite{gurobi} can be used for larger problem sizes.  To summarize, solving~\eref{eq:getM} can be done in $\CO(m p^n)$ flops, \emph{i.e.}, the complexity is linear with respect to the number of measurement locations and exponential in the number of strains.

The solution to \eref{eq:getMk} can be non-unique for two reasons.
First, it may be possible to find two different blocks $\bfM_k, \bfM_k' \in \tilde{\Omega}_M$, both minimizing \eref{eq:getMk}, such that $\bfM_k \bfw^{i-1} = \bfM_k' \bfw^{i-1}$, or equivalently $(\bfM_k-\bfM_k')\bfw^{i-1} = \bfzero$.
This leads to the definition of \emph{bi-independency} \cite{li2003}: The values in $\bfw$ are bi-independent if $\bfc^\top \bfw \neq 0$ for all $\bfc \in \{0,-1,1\}^n \setminus \{\bfzero\}$.
Clearly, for almost every $\bfw \in \Omega_W$ the values are bi-independent.

Second, there may be two different minimizing blocks $\bfM_k, \bfM_k' \in \tilde{\Omega}_M$ such that $\bfM_k \bfw^{i-1} \neq \bfM_k' \bfw^{i-1}$.
If $p=2$, this means that $\bfM_k \bfw^{i-1} - \bfd_k = \bfd_k - \bfM_k' \bfw^{i-1}$, or equivalently
\begin{equation} \label{eq:Mksum}
(\bfM_k+\bfM_k')\bfw^{i-1} = 2 \bfd_k.
\end{equation}
For $\bfd_k = 1/2$ this holds for every $\bfw^{i-1} \in \Omega_W$ if we choose $\bfM_k' = \bfone^\top - \bfM_k$.
Otherwise, for~\eref{eq:Mksum} to hold there must exist $\bfc \in \{0,1/2,1\}^n$ such that $\bfc^\top \bfw^{i-1} = \bfd_k$.
If $p>2$, the argument is not valid as such, but the general idea is still the same.
We conclude that~\eref{eq:getMk} has a unique solution for almost every $(\bfw^{i-1}, \bfd_k) \in \Omega_W \times \R^{p-1}$, and naturally~\eref{eq:getM} inherits a similar property as well.

In the second step of the block coordinate descent, we keep the binary matrix fixed and solve the convex quadratic programming problem~\eref{eq:getW} for the frequency vector.
Due to the equality constraint for $\bfw$, the solution for~\eref{eq:getW} is unique if the rank of $\bfM^{i}$ is at least $n-1$.
We note that the gradient and Hessian of the objective function $\phi$ with respect to the continuous variable $\bfw$ are
\begin{equation*}
    \nabla_{\bfw} \phi(\bfM,\bfw)  = \bfM^\top \Gamma^{-1} \left( \bfM \bfw - \bfd \right)
\end{equation*}
and
\begin{equation*}
    \nabla^2_{\bfw} \phi(\bfM,\bfw)  =  \bfM^\top \Gamma^{-1} \bfM,
\end{equation*}
respectively.
The update $\delta \bfw$ is then obtained by approximately solving the convex quadratic program 
\begin{eqnarray}\label{eq:ASQP}
    \min_{\delta\bfw} \quad & \hf \delta\bfw^\top \nabla^2_\bfw \phi(\bfM^{i},\bfw^{i-1}) \delta\bfw - \delta\bfw^\top \nabla_\bfw \phi(\bfM^{i},\bfw^{i-1}) \\
    \text{subject to} \quad & 0 \leq \bfw^{i-1} + \delta \bfw \leq 1,\; \sum_{j=1}^n \delta\bfw_j = 0. \nonumber
\end{eqnarray}
To this end, we use a few steps of a standard active set method for quadratic programming; see, \emph{e.g.},~\cite[Ch.~16]{NocedalWright2006} for a detailed description.

The block coordinate descent approach for MAP estimation is listed in Algorithm~\ref{alg:MAP}.
Since there are only finitely many instances of the problem~\eref{eq:getW}, and the value of the objective function cannot increase during the iteration, at some point the objective value must stagnate \cite{talwar1996}.
Thus, we repeat the steps~\eref{eq:getM} and~\eref{eq:getW} until there is no change in subsequent iterates $\bfM^{i}$ and $\bfM^{i-1}$.
In practice, one may also want to monitor the change of the frequency vector $\|\bfw^{i} - \bfw^{i-1}\|$ and set a maximum number for the iterations to make sure that the algorithm also stops in the case of non-unique solutions for the subproblems.

There is no guarantee that the iteration converges to a global minimum, which is why the block coordinate method is run $n_T \in \N$ times with different random initial vectors $\bfw^0$, and the output with the highest probability is selected as the MAP estimate.
This is usually a good strategy \cite{talwar1994,talwar1996,slawski2013}; see also \cite{ChanWong2000}, where the dependency of the solution on the starting guess was established for the blind deconvolution problem without binary constraints.

In addition to the block-wise non-uniqueness stemming from the subproblems \eref{eq:getM}--\eref{eq:getW}, the global minimum may be obtained with multiple elements of $\Omega := \Omega_M \times \Omega_W$ such that both blocks, \emph{i.e.}, the binary matrices and frequency vectors, differ.
This can be interpreted as a generalization of the permutation invariance which is eliminated by our choice for the prior: If $\bfQ \in \R^{n \times n}$ is a matrix such that $\bfM \bfQ \in \Omega_M$ and $\bfQ^{-1} \bfw \in \Omega_W$, then clearly $(\bfM,\bfw)$ and $(\bfM \bfQ, \bfQ^{-1} \bfw)$ correspond to the same value of the objective function $\phi$.
The existence of such nontrivial $\bfQ$ is discussed in detail in~\cite{behr2017}.
In short, the uniqueness (up to permutations) of the factorization $\bfM \bfw$ becomes rapidly more likely when the number of distinct rows in $\bfM$ increases.

\begin{algorithm}[t]
\caption{Block coordinate descent for strain identification from mixed samples.}
\label{alg:MAP}
\begin{algorithmic}
\STATE \textbf{Input:} Measurements $\bfd \in \R^q$, number of strains $n \in \N$, number of classes $p \geq 2$, number of trials $n_T \in\N$, tolerance $\eps_w > 0$, maximum number of iterations $n_I \in \N$.
\FOR{$t=1,2,\ldots,n_T$}
    \STATE Draw starting guess $\bfw^0$ uniformly from $\Omega_W$ and set, \emph{e.g.}, $\bfM^0 = -\bfone$.
    \FOR {$i=1,\ldots,n_I$}
        \STATE Get $\bfM^{i}$ block-wise by solving~\eref{eq:getMk} for current $\bfw^{i-1}$.
        \STATE Get $\bfw^{i}$ by solving~\eref{eq:ASQP} for current $\bfM^{i}$.
        \IF {$\bfM^{i} = \bfM^{i-1}$ and $\| \bfw^{i} - \bfw^{i-1} \| < \eps_w$}
            \STATE Exit the inner loop.
        \ENDIF
     \ENDFOR
    \STATE Store local mode: $(\hat{\bfM}^{t},\hat{\bfw}^t) = (\bfM^{i},\bfw^{i})$.
\ENDFOR
\STATE Find the best mode: $\ell = \argmin_{1 \leq t \leq n_T} \phi(\hat{\bfM^t},\hat{\bfw}^t)$
\STATE \textbf{Output:} MAP estimate $(\hat{\bfM},\hat{\bfw}) = (\hat{\bfM^\ell},\hat{\bfw}^\ell)$ or (if desired) all modes $(\hat{\bfM}^1,\hat{\bfw}^1),\ldots,(\hat{\bfM}^{n_T},\hat{\bfw}^{n_T})$.
\end{algorithmic}
\end{algorithm}

The prior distributions for both $M$ and $W$ involve the knowledge about the number of strains $n$, \emph{i.e.}, the multiplicity of infection.
As mentioned, this number may be unknown in many practical applications.
An alternative to the method described above is to resort to the discrepancy principle with an approach that resembles the so-called ``regularization by discretization'' technique~\cite{Hamarik2016,Hansen2010}.
To this end, let $(\hat{\bfM}(n), \hat{\bfw}(n))$ denote the MAP estimate for a given $n \geq 1$.
Now the goal is to find $n$ such that the discrepancy between the measurement and the reconstruction is approximately equal to the magnitude of noise, which is still assumed to be known.
More precisely, we start from $n=1$ and keep increasing $n$ until
\begin{equation*}
    d(n) := \| \hat{\bfM}(n) \hat{\bfw}(n) - \bfd \|_2^2 \leq \sum_{i=1}^q \gamma_i^2.
\end{equation*}
Note that $d$ is a non-increasing function if the MAP estimates are global minimizers of the objective $\phi$.

\subsection{MAP estimation as a convex mixed-integer quadratic program}
\label{sub:miqp}

Although the block coordinate descent method is computationally simple and efficient in finding local minima, there is no guarantee that it yields a global minimum.
Even with numerous trials, one may end up finding only local minima.
However, we can reformulate the problem as a \emph{mixed-integer quadratic program} (MIQP) with a convex objective function and linear constraints in addition to the binary restriction on the matrix $\bfM$. For moderate-sized instances, this class of program can be efficiently solved to global optimality by a commercial off-the-shelf solver such as Gurobi~\cite{gurobi} or CPLEX~\cite{cplex}.

In short, we reformulate the problem by replacing the bilinear term $\bfM_{i,j} \bfw_j$ with its so-called McCormick envelope for $i=1,\ldots,q$ and $j=1,\ldots,n$. We then use the fact that $\bfM$ is binary and every component of $\bfw$ is bounded between $0$ and $1$ to prove the equivalence between the two formulations. See \cite{Shabbir2013} for a general treatment of this technique and \cite{Gupte2017,Schumacher2017} for two examples of applications.

To write the McCormick envelopes, we define the auxiliary variables
\begin{equation*}
    \bfZ_{i,j} := \bfM_{i,j} \bfw_j, \qquad i=1,\ldots,q, \quad j=1,\ldots,n.
\end{equation*}
Using these new variables, problem~\eref{eq:MAP} can be equivalently written as
\begin{eqnarray}
    \min  &&\sum_{i=1}^q \frac{1}{\gamma_i^2} \left( \sum_{j=1}^n \bfZ_{i,j} - \bfd_i\right)^2 \label{eq:MAP_z}\\
\text{subject to} \quad &&\bfZ_{i,j} = \bfM_{i,j} \bfw_j, \qquad i=1,\ldots,q, \quad j=1,\ldots,n,\nonumber\\
&&\bfM \in \Omega_M,\nonumber\\
&&\bfw \in \Omega_W.\nonumber
\end{eqnarray}
Notice that in~\eref{eq:MAP_z} the objective is convex and all the non-convexity comes from the bilinear constraints defining $\bfZ$. Next, we replace each bilinear constraint by a convex envelope given by the McCormick's inequalities \cite{McCormick1976}:
\begin{eqnarray}
    \min  &&\sum_{i=1}^q \frac{1}{\gamma_i^2} \left( \sum_{j=1}^n \bfZ_{i,j} - \bfd_i\right)^2 \label{eq:MAP_z_mc}\\
\text{subject to} \quad &&\bfZ_{i,j} \geq 0,\nonumber\\
&&\bfZ_{i,j} \leq \bfM_{i,j},\nonumber\\
&&\bfZ_{i,j} \leq \bfw_j,\nonumber\\
&&\bfZ_{i,j} \geq \bfM_{i,j} + \bfw_j - 1, \qquad i=1,\ldots,q, \quad j=1,\ldots,n,\nonumber\\
&&\bfM \in \Omega_M,\nonumber\\
&&\bfw \in \Omega_W.\nonumber
\end{eqnarray}
We claim that the problems~\eref{eq:MAP_z} and~\eref{eq:MAP_z_mc} are equivalent. Since both problems have the same objective, it is enough to show that both problems have the same set of feasible solutions. Let $S_1$ and $S_2$ denote the set of feasible solutions of problem~\eref{eq:MAP_z} and problem~\eref{eq:MAP_z_mc}, respectively. By construction, $S_1\subseteq S_2$. Next we show that $S_1\supseteq S_2$. Let $(\bfM,\bfw,\bfZ)\in S_2$. Then, for all $i=1,\ldots,q$ and $j=1,\ldots,n$, there are two cases:
\begin{enumerate}
\item[(i)] $\bfM_{i,j}=0$: In this case, the McCormick's inequalities imply that $\bfZ_{i,j} = 0$ and $0\leq\bfw_j\leq 1$. Thus, $\bfZ_{i,j} = \bfM_{i,j} \bfw_j$ in this case.
\item[(ii)] $\bfM_{i,j}=1$: In this case, the McCormick's inequalities imply that $0\leq\bfZ_{i,j}\leq 1$ and $\bfZ_{i,j} = \bfw_j$. Thus, $\bfZ_{i,j} = \bfM_{i,j} \bfw_j$ in this case as well.
\end{enumerate}
Therefore, $(\bfM,\bfw,\bfZ)\in S_1$, and we conclude that $S_1=S_2$ and hence the problems are equivalent.

Following, \emph{e.g.}, a branch-and-bound strategy, we fix some of the binary-constrained entries $\bfM_{i,j}$ to be either $0$ or $1$, and relax the binary restrictions from the remaining entries (see, \emph{e.g.}, \cite{BelottiEtAl2013} for a general overview of MINLP solvers). 
This way~\eref{eq:MAP_z_mc} becomes a convex quadratic program with linear constraints and continuous variables.
Solving this relaxed problem is straightforward and similar to~\eref{eq:ASQP}.
By alternating the fixed entries in a tree-like fashion and comparing the minima of the relaxed problems, we establish upper and lower bounds for the problem~\eref{eq:MAP_z_mc} and thus for the minimum of the original target function $\phi$.
This tree is traversed until a desired gap between the upper and lower bounds is achieved.

\subsection{Integrating the posterior density}
\label{sub:integration}

Next, we discuss how to compute integrals that include the posterior density \eref{eq:posterior}.
This becomes useful when one wants to compute conditional moments such as mean or variance of the random variables $M$ and $W$, given the observations $D$.

Let $f$ be a function that satisfies
\begin{equation} \label{eq:f}
    f(\bfM,\bfw) = \sum_{k=1}^m f_k(\bfM_k, \bfw),
\end{equation}
that is, $f_k$ depends only on the $k$th block of $\bfM$ in addition to $\bfw$.
Examples of such functions include $f(\bfM,\bfw) = \bfM$ and $f(\bfM,\bfw) = \bfw$, as well as the entrywise powers of $\bfM$ and $\bfw$.
Here we present an integration scheme for computing the posterior mean of $f$.
More precisely, we consider the integral
\begin{eqnarray}
    \E[f(M,W) \mid D] &= \int_\Omega f(\bfM,\bfw) \pi(\bfM,\bfw \mid \bfd) \,\mathrm{d}\bfM \mathrm{d}\bfw \nonumber \\
    &= \int_{\Omega_W} \sum_{\bfM \in \Omega_M} f(\bfM,\bfw) \pi(\bfM,\bfw \mid \bfd) \,\mathrm{d}\bfw. \label{eq:postint2}
\end{eqnarray}
Na\"ively summing over all possible binary matrices is impractical even for moderate parameter values, since $|\Omega_M| = p^{mn}$.
Thus, we suggest a more efficient approach that exploits the separability of the posterior in the same fashion as \eref{eq:getMk}.

First, note that the posterior density can be written as
\begin{equation} \label{eq:postprod}
    \pi(\bfM,\bfw \mid \bfd) = C \exp \mathopen{}\left( \sum_{i=1}^m g_i(\bfM_i) \right)\mathclose{} 
    = C \prod_{i=1}^m \exp \mathopen{}\left( g_i(\bfM_i) \right)\mathclose{},
\end{equation}
where $C$ is a constant that depends on $\bfd$ and that can be computed by considering the case $f=1$, and
\begin{equation*}
    g_i(\bfM_i) := -\hf \left\| \bfM_{i} \bfw - \bfd_i\right\|^2_{\Gamma_i}.
\end{equation*}
Next, the sum in \eref{eq:postint2} can be decomposed by iterating through the blocks separately, that is,
\begin{equation*}
    \sum_{\bfM \in \Omega_M} f(\bfM,\bfw) \pi(\bfM,\bfw \mid \bfd) = \sum_{\bfM_1 \in \tilde{\Omega}_M} \cdots \sum_{\bfM_m \in \tilde{\Omega}_M} f(\bfM,\bfw) \pi(\bfM,\bfw \mid \bfd).
\end{equation*}
By substituting the expressions \eref{eq:f} and \eref{eq:postprod} into this sum, we can write the integrand in \eref{eq:postint2} as
\begin{eqnarray}
    & C \sum_{\bfM_1 \in \tilde{\Omega}_M} \cdots \sum_{\bfM_m \in \tilde{\Omega}_M} \prod_{i=1}^m \exp \mathopen{}\left( -\hf g_i(\bfM_i) \right)\mathclose{} \sum_{k=1}^m f_k(\bfM_k,\bfw) \nonumber\\
    =&C \sum_{k=1}^m \sum_{\bfM_1 \in \tilde{\Omega}_M} \exp \mathopen{}\left(g_1(\bfM_1) \right)\mathclose{} \ \cdots \sum_{\bfM_m \in \tilde{\Omega}_M} \exp \mathopen{}\left( g_m(\bfM_m) \right)\mathclose{} f_k(\bfM_k,\bfw) \label{eq:inttech}
\end{eqnarray}
For each $k$ in the outermost sum, $f_k$ can be put inside the sum that corresponds to the $k$th block of the binary matrix.
As a result, the integrand in \eref{eq:postint2} becomes a sum where each summand is a product of sums.

The rapid decay of the exponential function can easily trigger underflows when using floating point arithmetics.
Therefore, a numerically more stable version of the integration technique is outlined in Algorithm~\ref{alg:integration}.
We see that for each quadrature node $\bfw \in \Omega_W$, the evaluation of the integrand has a computational complexity of $\CO(mp^n)$.
For the integral over $\Omega_W$, we could apply some deterministic quadrature rule if $n$ is small, but in the numerical experiments in the next section, we will instead use Monte Carlo method for simplicity~\cite{JacodProtter2004}.

\begin{algorithm}[t]
\caption{Computing the conditional mean of a function $f$.}
\label{alg:integration}
\begin{algorithmic}
\STATE \textbf{Input:} Measurements $\bfd \in \R^q$, function $f$ of the separable form \eref{eq:f}, quadrature nodes and weights $\{\bfw^{(s)}, \zeta^{(s)}\}_{s=1}^S \subset \Omega_W \times \R$, number of strains $n$, number of measurement locations $m$, number of classes $p$.
\FOR{$s=1,\ldots,S$}
    \FOR{$k=1,\ldots,m$}
        \FOR{$j=1,\ldots,p^n$}
            \STATE Let $\bfM^{(j)}$ be the $j$th element of $\tilde{\Omega}_M$ (in some arbitrary but fixed order)
            \STATE $L_{k,j} = -\hf \| \bfM^{(j)} \bfw^{(s)} - \bfd_k \|^2_{\Gamma_k}$
            \STATE $F_{k,j} = f_k(\bfM^{(j)}, \bfw^{(s)})$
        \ENDFOR
        \STATE $U_k = \max_j L_{k,j}$ (for numerical stability)
        \STATE $P_k = \sum_j \exp(L_{k,j} - U_k)$
        \STATE $G_k = \sum_j F_{k,j} \exp(L_{k,j} - U_k)$
    \ENDFOR
    \STATE Compute sum \eref{eq:inttech}: $J_s(f) = G_1 P_2 P_3 \cdots P_m + P_1 G_2 P_3 \cdots P_m + \ldots + P_1 \cdots P_{m-1} G_m$
    \STATE $J_s(1) = \prod_{k=1}^m P_k$ (corresponds to $f = 1$)
    \STATE $\lambda_s = \sum_{k=1}^m U_k$
\ENDFOR
\STATE Compute unnormalized integrals with the quadrature rule:
\STATE $I(f) = \sum_s \zeta_s J_s(f) \exp(\lambda_s - \max_\ell \lambda_\ell)$
\STATE $I(1) = \sum_s \zeta_s J_s(1) \exp(\lambda_s - \max_\ell \lambda_\ell)$
\STATE \textbf{Output:} Posterior mean of $f$ as $I(f) / I(1)$.
\end{algorithmic}
\end{algorithm}

\section{Numerical experiments}
\label{sec:experiments}
We validate our computational techniques using both synthetic and real data with known ground truths.
In Section~\ref{sub:ex1} we illustrate the problem of non-uniqueness with some simple examples.
We then introduce more realistic reconstruction problems in Section~\ref{sub:exmap} and study how different parameter values affect the accuracy of the reconstruction.
Finally, in Section~\ref{sub:exuq} we apply our methods to experimental data.

\subsection{Implementation}
\label{sub:implementation}
Our computational experiments are performed using the Julia~\cite{BezansonEtAl2012} programming environment. 
We have implemented a module for strain identification that includes the block coordinate descent method and the numerical integration technique. For the convex problem formulation, we use Gurobi~\cite{Dunning2015} through an interface of the Julia module JuMP to solve the problem with a global optimization method. Our implementation is freely available at
\begin{center}
	\url{https://github.com/lruthotto/StrainRecon.jl/}
\end{center}

\subsection{Identifiability}
\label{sub:ex1}
In this section we demonstrate with a few simple examples that the solution to the MAP estimation problem can be either unique or non-unique.
The examples are chosen to be small such that full enumeration of the binary matrices are possible.
That is, the results can be confirmed by solving $|\Omega_M| = p^{mn}$ quadratic programming problems \eref{eq:ASQP}.
For larger problems this would not be feasible.
Therefore, we also illustrate how the possible ambiguity can be seen in the posterior statistics.

First, let us choose $m=3$ and $n=p=2$, and consider the following example that illustrates the concept of bi-independency from Section~\ref{sub:map_estimation}.
Let $\bfw^{(1)} = (0.6,0.4)^\top$, $\bfw^{(2)} = (0.5,0.5)^\top$, and let $\bfM \in \{0,1\}^{3\times 2}$ be
	\begin{equation*}
		\bfM = \left( 
		\begin{array}{rrr}
		 	0 & 1 \\
		 	1 & 0 \\
		 	1 & 1 
		\end{array}
		\right).
	\end{equation*}
	Note that $\bfc^\top \bfw^{(2)} = 0$ for $\bfc = (1, -1)$.
	The data is given by
	\begin{equation*}
		\bfd^{(1)} = \bfM \bfw^{(1)} = \left( \begin{array}{r}
			0.4 \\
		 	0.6 \\
		 	1.0 \\
		\end{array}\right)
		\quad \text{ and } \quad
		\bfd^{(2)} = \bfM \bfw^{(2)} = \left( \begin{array}{r}
			0.5 \\
		 	0.5 \\
		 	1.0 \\
		\end{array}\right).
	\end{equation*}
	One can readily verify that there are no other pairs in $\Omega$ that would yield the first data vector $\bfd^{(1)}$, therefore the inverse problem has a unique solution.
	In contrast, $\bfd^{(2)}$ can be obtained by choosing $\bfw^{(2)}$ as above and any matrix in $\Omega_M$ which has row sums of $1$ for the first two rows and $2$ for the third row.
	Obviously, there are four such matrices.
	
We use the integration technique presented in Section~\ref{sub:integration} to compute the posterior mean (\emph{i.e.}, conditional mean) and standard deviation for the unknowns $\bfM$ and $\bfw$.
Throughout this section, we assume that the noise covariance matrix in \eref{eq:posterior} is $\Gamma = \gamma^2 I$ for some standard deviation $\gamma>0$.
For large values of $\gamma$ we expect to see larger standard deviation in the posterior and the posterior mean is expected to approach the mean of the prior.
However, already with $\gamma = 10^{-2}$ the posterior means $\bfM_{\rm CM}^{(1)}$ and $\bfw_{\rm CM}^{(1)}$ corresponding to $\bfd^{(1)}$ are practically indistinguishable from the true values and the posterior variance of $\bfM_{\rm std}^{(1)}$ is numerically zero.
On the other hand, the standard deviation $\bfw_{\rm std}^{(1)}$ of the frequency vector is approximately $(0.007, 0.007)^\top$.
As a comparison, for $n=2$, the standard deviation of the essentially $1$-dimensional uniform distribution on $\Omega_W$ is approximately $0.14$.

As already seen above, $\bfd^{(2)}$ has more uncertainty in the reconstruction.
This can also be verified by computing the posterior moments, since now we obtain
\begin{equation*}
	\bfM_{\rm CM}^{(2)} \approx \left(\begin{array}{rr}
			0.5 & 0.5 \\
		 	0.5 & 0.5 \\
		 	1 & 1 \\
		\end{array}\right)
		\quad \text{ and } \quad
		\bfM_{\rm std}^{(2)} \approx \left(\begin{array}{rr}
			0.5 & 0.5 \\
		 	0.5 & 0.5 \\
		 	0 & 0 \\
		\end{array}\right),
\end{equation*}
which are in line with the earlier observations.
The frequency vector, however, has less uncertainty in the second case.
The posterior mean is close to $(0.5, 0.5)^\top$, as expected, and the standard deviation is only $\bfw_{\rm std}^{(2)} \approx (0.004, 0.004)^\top$.

\begin{figure}[t]
	\begin{center}
		\includegraphics{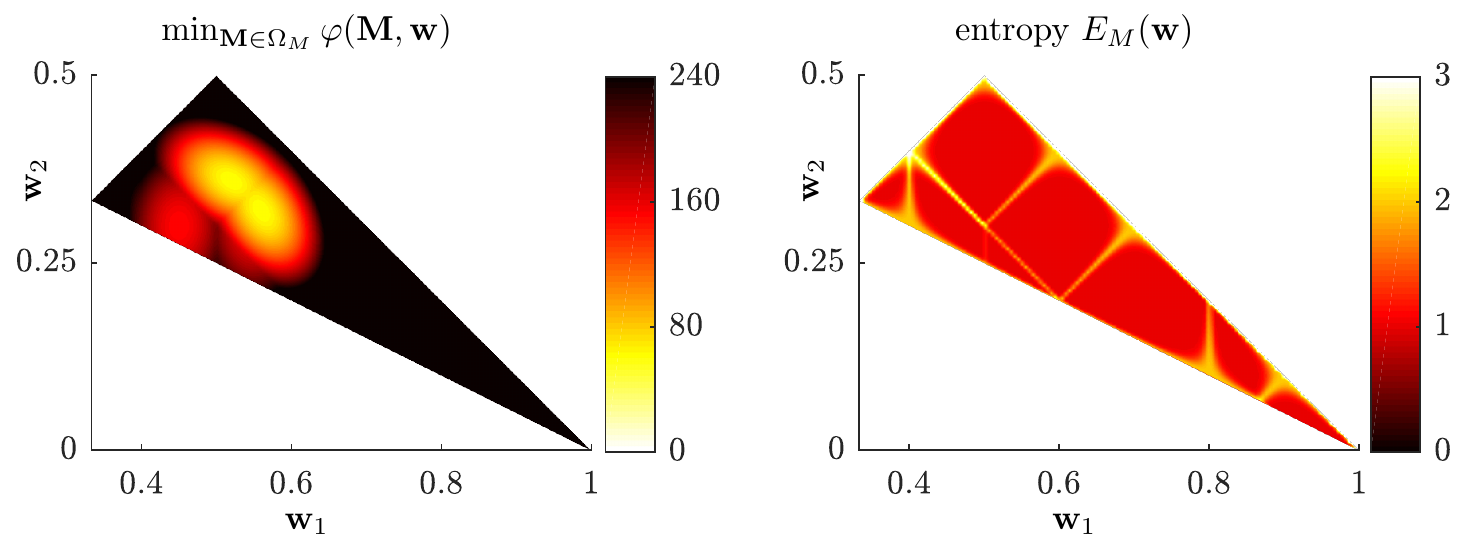}
	\end{center}
	\caption{\textbf{Left:} Minimum target function value as a function of the frequency vector $\bfw$ for the data $\bfd$ given in \eref{eq:data4}. \textbf{Right:} Entropy of $\pi(\bfM \mid \bfw, \bfd)$ for the same example, computed by \eref{eq:entropy}.}
	\label{fig:4by3}
\end{figure}

As a next example, we consider the identification problem for $(m,n,p) = (4,3,2)$ from the data 
\begin{equation}\label{eq:data4}
	\bfd = (0.1, 0.3, 0.5, 0.6)^\top.
\end{equation}
No exact solution for the inverse problem can be found, but four global minima for~\eref{eq:MAP} can be obtained.
These correspond to two different frequency vectors and four different binary matrices.
The left side of Figure~\ref{fig:4by3} shows $\min_{\bfM \in \Omega_M} \varphi(\bfM,\bfw)$ for different frequency vectors with $\gamma = 10^{-2}$.
The global minima can be seen at $\bfw = (0.52, 0.36, 0.12)^\top$ and $\bfw = (0.56, 0.32, 0.12)^\top$.
In addition, there is at least one local minimum at $\bfw = (0.45, 0.30, 0.25)^\top$, which we occasionally obtain as the output of our block coordinate descent algorithm.

The posterior mean and standard deviation for the binary matrix $\bfM$ are
\begin{equation*}
	\bfM_{\rm CM} \approx \left(\begin{array}{rrr}
			0 & 0 & 1 \\
		 	0 & 1 & 0 \\
		 	0.5 & 0.5 & 0.5 \\
		 	1 & 0 & 0.5 \\
		\end{array}\right)
		\quad \text{ and } \quad
		\bfM_{\rm std} \approx \left(\begin{array}{rrr}
			0 & 0 & 0 \\
		 	0 & 0 & 0 \\
		 	0.5 & 0.5 & 0.5 \\
		 	0 & 0 & 0.5 \\
		\end{array}\right),
\end{equation*}
respectively.
The uncertainty in the third row results from having the value $0.5$ in $\bfd$ (see Section~\ref{sub:map_estimation}), whereas two different values in the lower right corner of $\bfM$ correspond to two different frequency vectors.
For the frequency vector we obtain $\bfw_{\rm CM} \approx (0.54, 0.34, 0.12)^\top$ and $\bfw_{\rm std} \approx (0.022, 0.021, 0.008)^\top$.

The right hand side of Figure~\ref{fig:4by3} shows the entropy of the distribution $\pi(\bfM \mid \bfw,\bfd)$, defined as a function of the frequency vector $\bfw$
\begin{equation}\label{eq:entropy}
	E_M(\bfw) := \sum_{\bfM \in \Omega_M} \pi(\bfM \mid \bfw,\bfd) \log_2\Big(\pi(\bfM \mid \bfw,\bfd)\Big).
\end{equation}
The entropy clearly indicates areas of $\Omega_W$ where the matrix minimizer of $\varphi(\bfM,\bfw)$ is highly non-unique.
For example, for $\bfw = (0.5, 0.3, 0.2)^\top$ there are $12$ binary matrices $\bfM$ that result in the same minimal value of $\varphi(\bfM,\bfw)$.
Note that the entropy \eref{eq:entropy} can be efficiently computed also for larger examples by using the same row-decoupling technique as in \eref{eq:getMk}.

Finally, let us mention that both MAP estimation techniques introduced in Sections \ref{sub:map_estimation} and \ref{sub:miqp} reliably find a global minimizer in all previous examples, except in the last example where a local minimizer is sometimes returned by the block coordinate descent method if $n_T$ is small.
Which of the global minimizers is found depends on the starting points $\bfw^0$ and the implementation details; for example, how the minimizing matrix in~\eref{eq:getM} is chosen in case it is not unique.

\subsection{Accuracy of the MAP estimation}
\label{sub:exmap}
We validate the accuracy of the block coordinate descent method (see Algorithm~\ref{alg:MAP}) and the convex MIQP formulation (see Section~\ref{sub:miqp}) using $10\,000$ randomly generated data sets with different noise levels.

Before quantifying the accuracies of our reconstruction methods, we have to define a meaningful distance function between the ground truth $(\bfM,\bfw)$ and the reconstruction $(\hat{\bfM},\hat{\bfw})$, or more generally, a distance between any two pairs in $\Omega$.
We first note that every misclassification in the binary matrix should have equal impact on the distance, that is, it should not matter whether we identify the first class as the third class or the third class as the second class, and so on.
To ensure this property holds true, both binary matrices are augmented to matrices in $\{0,1\}^{mp \times n}$ by adding the missing first row of each block so that the column sums of each block become exactly $1$.
We denote this modification of a matrix $\bfM$ by $\tau(\bfM)$.
Another observation is that the order of the strains does not matter; the requirement that the frequencies are in non-increasing order is for computational purposes only.
As an extreme example, if we had
\begin{equation*}
	(\bfM,\bfw) = \left(\left(
		\begin{array}{rrr}
			1 & 0 \\
			1 & 0
		\end{array}
	\right), \left(
		\begin{array}{rrr}
			0.51 \\
			0.49
		\end{array}
	\right)\right), \quad
	(\hat{\bfM},\hat{\bfw}) = \left(\left(
		\begin{array}{rrr}
			0 & 1 \\
			0 & 1
		\end{array}
	\right), \left(
		\begin{array}{rrr}
			0.51 \\
			0.49
		\end{array}
	\right)\right),
\end{equation*}
the strains would be identified perfectly and their frequencies would be reconstructed accurately, so we would expect a small distance.
Therefore, we minimize over all possible permutations of the strains before computing the distance.

The distance, or reconstruction error $e$, can now be defined as
\begin{equation} \label{eq:recoerr}
	e(\bfM,\bfw,\hat{\bfM},\hat{\bfw}) := \min_{\bfP \in \sigma(n)} \| \tau(\bfM) \diag(\bfw) - \tau(\hat{\bfM}) \diag(\hat{\bfw}) \bfP \|_1,
\end{equation}
where $\sigma(n)$ is the set of permutation matrices of size $n \times n$ and $\| \cdot \|_1$ denotes the entrywise $1$-norm, \emph{i.e.}, the usual $\ell^1$-norm after vectorization.
Enumerating all permutations is feasible in our examples due to the small number of strains, $n$.

Let us study the distribution of the reconstruction error by sampling realizations $(\bfM,\bfw) \in \Omega$ and computing the corresponding MAP estimates $(\hat{\bfM},\hat{\bfw})$ with Algorithm~\ref{alg:MAP}, where $n_T=20$, $\eps_w = 10^{-3}$ and $n_I=10$.
For given values of $m$, $n$, and $p$, we draw $10\,000$ independent samples from the uniform prior distribution $\pi(\bfM,\bfw)$ with the additional restriction that the matrix $\bfM$ must not contain duplicate columns.
Before computing the MAP estimate, the data $\bfd = \bfM\bfw$ is contaminated with independent zero mean Gaussian noise with standard deviation $\gamma>0$.
For comparison, the reconstructions for the same noisy data are also computed after converting the objective function to convex form as described in Section~\ref{sub:miqp}.
The resulting MIQP problems are solved using Gurobi software with ``MIPGap'' tolerance parameter set to $10^{-6}$.

\begin{figure}[t]
	\begin{center}
		\includegraphics{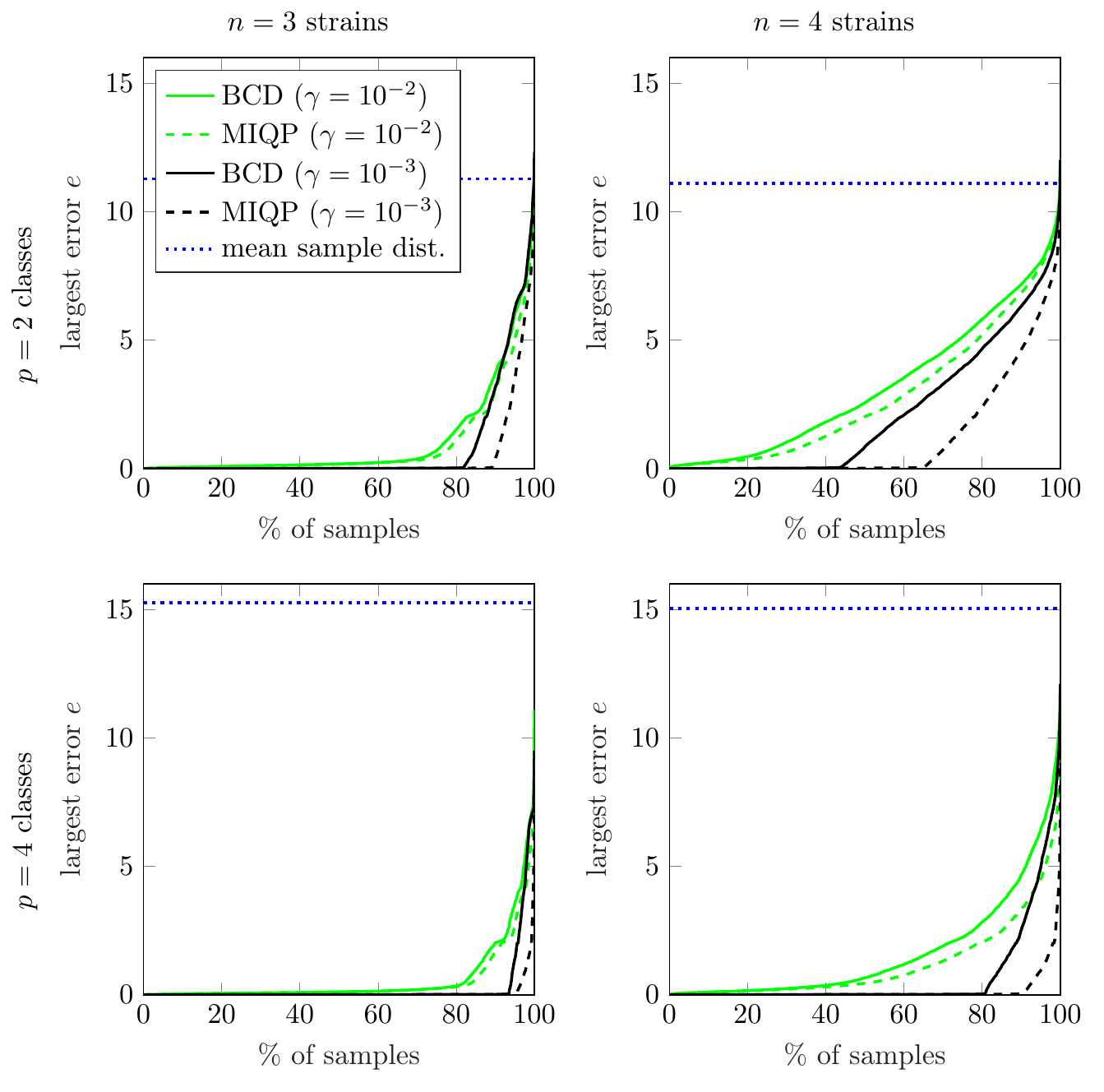}
	\end{center}
	\caption{Sorted reconstruction errors for $m=10$ measurement sites and Gaussian measurement error with standard deviation $\gamma=10^{-2}$ (green) and $\gamma=10^{-3}$ (black). The solid line depicts the block coordinates descent; the dashed line corresponds to the convex MIQP. The horizontal dashed line shows the average distance between the random samples.}
	\label{fig:errors}
\end{figure}

Figure~\ref{fig:errors} shows the reconstruction errors $e$ for $m =10$, $n \in \{3,4 \}$, $p\in\{2,4 \}$ and $\gamma \in \{10^{-2},10^{-3}\}$.
For clarity, all reconstruction errors are sorted in ascending order.
The average distance $e$ between two random samples is shown for each case by a horizontal line.
The first observation is that both reconstruction methods perform significantly better than just randomly drawing the reconstruction, even with uniform priors.
We also notice that the reconstruction error increases when the number of strains, $n$, is increased, but decreases when the number of classes, $p$, is increased.

Comparing the two MAP estimation methods, we see that solving the convex MIQP problem yields smaller statistical error in all cases, compared to the block coordinate descent.
For example, when $(n,p) = (4,2)$, the former produces negligible error in almost two thirds of the samples, whereas for the latter, fewer than half of the samples are reconstructed with such accuracy.
However, it is expected that increasing the number of trials, $n_T$, in Algorithm~\ref{alg:MAP} would make the block coordinate descent method perform better.

Unsurprisingly, the reconstruction errors become larger when the measurement noise is increased.
In addition, with $\gamma = 10^{-2}$ the difference between the two reconstruction methods is less evident than with the smaller noise level.

\subsection{Experimental data and uncertainty quantification}
\label{sub:exuq}

The initial motivation for the strain reconstruction was to tackle the practical challenge of disambiguating malarial strains.
We now apply our algorithms to the open experimental dataset previously analyzed by Zhu \emph{et al.}~\cite{Zhu2018}.
The dataset is generated from lab-mixed \emph{in vitro} samples of DNA from four laboratory parasite strains (3D7, Dd2, HB3, and 7G8) that are mixed in
27 different proportions. In a process similar to that of Section~\ref{sec:bg}, each of the 27 samples is sent to the MalariaGEN pipeline~\footnote{The Malaria Genomic Epidemiology Network: \url{http://www.malariagen.net/}} and genotyped with an Illumina sequencing platform to
produce a measurement vector.
Since the mixture proportions are controlled, the underlying ground truth strain barcodes and frequencies $(\bfM,\bfw)$ are known.

Only three of the 27 samples contained $n=3$ strains, specifically \mbox{PG0395-C}, \mbox{PG0396-C}, and \mbox{PG0397-C}, and the remaining samples contained either a single strain or two strains.
\mbox{PG0395-C} is a mixture of three parasite strains in near equal proportion, representing the edge case for identifiability where our algorithm has no basis for disambiguation (see discussion on bi-dependency in Section~\ref{sub:map_estimation}).

To illustrate the power of our algorithm in a challenging scenario, we focus on sample \mbox{PG0397-C} that contains $n=3$ strains in proportions 1:1:5.
We compute $\bfd$ for each of the $17\,420$ biallelic SNP sites of the sample by parsing the genomic sequences (reads) output by the Illumina sequencer in the VCF (Variant Call Format) format. 
At each SNP site, we used the proportion ${\text{alternate read}}/({\text{alternate read + reference read}})$ for the $\bfd$ vector component, where alternate read and reference read refer to the number of reads that support the alternate allele (``non-reference'') or reference allele present at the given SNP site, respectively.

Field samples often have low parasite copy numbers, and the number of SNP locations with recoverable allele frequencies will not necessarily reflect full genome coverage.
We choose to compare the 16 SNP sites recovered from the Daniels \emph{et~al.}~\cite{daniels2008} 24 SNP barcoding scheme in this sample set to allow direct evaluation of the two approaches.
In our algorithms, we thus set $m=16$ with $p=2$.

The sample standard deviation of the error $\bfd - \bfM\bfw$ in the data is about $0.05$.
In our experiment, we assume the noise vector $\bfn$ to be a Gaussian random variable with zero mean and standard deviation $\gamma=10^{-1}$.
While more elaborate noise models may be used in practice, our goal is to demonstrate that a simple noise model works sufficiently well with real data when the standard deviation parameter is chosen appropriately.

\begin{figure}[t]
\begin{center}
	\includegraphics{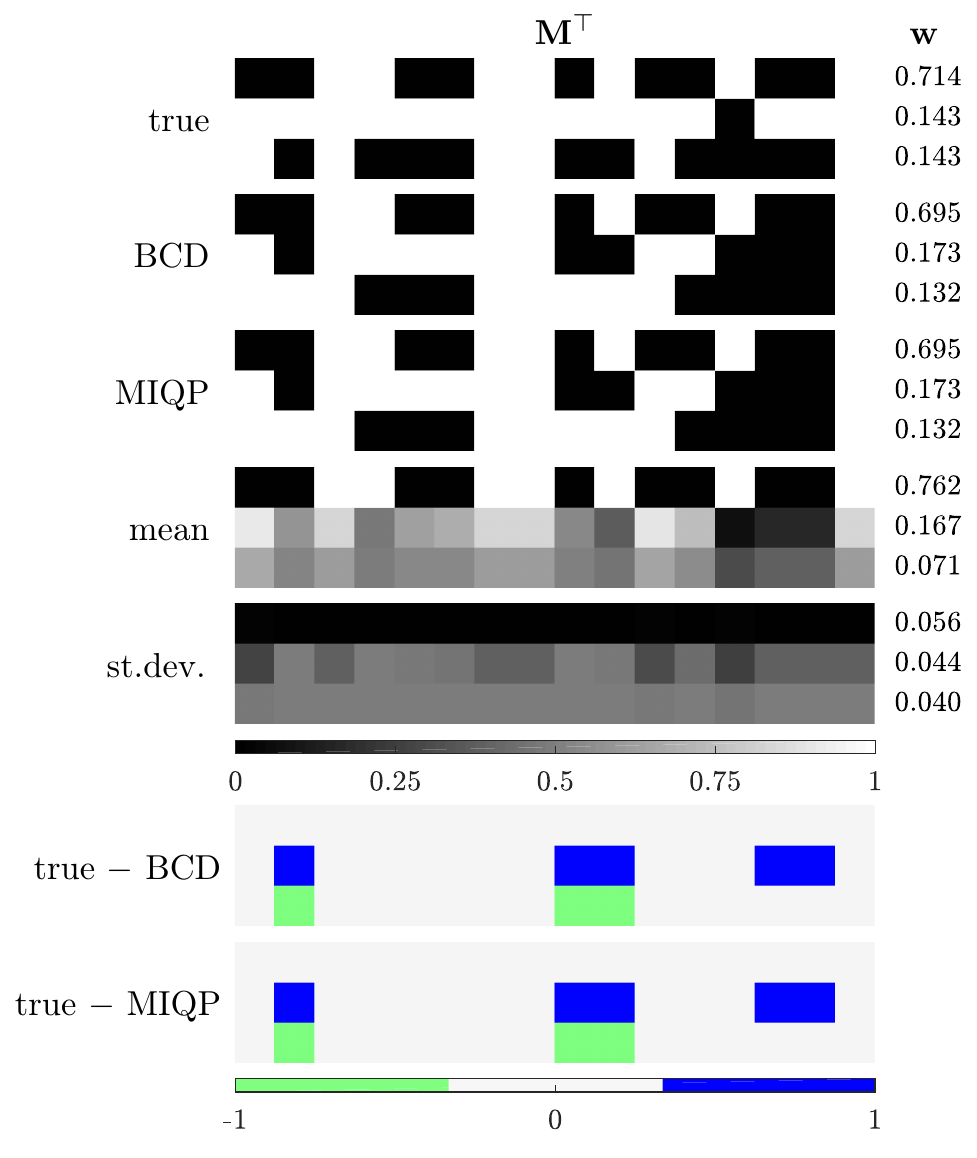}
	\caption{Ground truth and reconstructed $\bfM$ and $\bfw$ from using experimental measurement of $\bfd$ with $m=16$, $p=2$,  $n=3$ and assumed Gaussian random noise with zero mean and standard deviation $\gamma=10^{-1}$. The relative strain frequencies are sorted from highest to lowest and shown to the right of their corresponding SNP barcodes. From top to bottom: $\bfM$ and $\bfw$ corresponding to the ground truth, two MAP estimates from Sections \ref{sub:map_estimation} and \ref{sub:miqp}, and the conditional mean and standard deviation.}
	\label{fig:realexample}
\end{center}
\end{figure}

The strain reconstructions $\hat{\bfM}$ and $\hat{\bfw}$ from the block coordinate descent ($n_T=200$, $\eps_w = 10^{-3}$ and $n_I=10$) and  the convex MIQP formulation are identical, as shown in Figure~\ref{fig:realexample}.
The figure also shows the ground truth $(\bfM,\bfw)$, the conditional means and the posterior standard deviations.
As expected with a 1:1:5 mixture, we observe larger standard deviations and reconstruction errors in the SNP barcodes of the two less prominent strains using either of the MAP estimation methods.
This experiment also highlights the problem of identifiability mentioned in Sections~\ref{sub:map_estimation} and~\ref{sub:ex1}, as the two less prominent strains have an equal true relative frequency of $0.143$. 
In contrast, we see a perfect reconstruction of the SNP barcode associated with the most prominent strain which has a true relative frequency of $0.714$.

\begin{figure}[t]
	\begin{center}
		\includegraphics{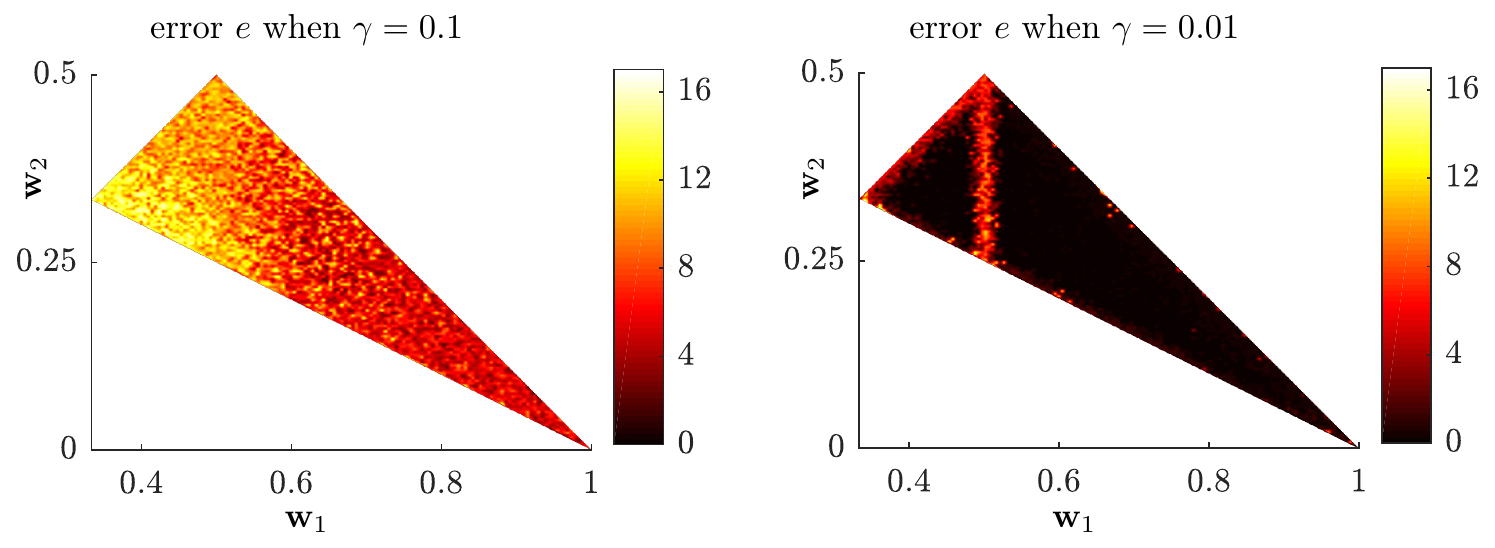}
	\end{center}
	\caption{Reconstruction error $e$ \eref{eq:recoerr} from block coordinate descent MAP estimation for varying $\bfw$ given a fixed $\bfM$ and two noise levels. For noise level $\gamma=10^{-1}$, higher reconstruction errors are observed in comparison to the same instances of $\bfw$ for the noise level $\gamma=10^{-2}$. Cases where a component of $\bfw$ can (approximately) be expressed as a sum of one or two other components are reflected in higher reconstruction errors for the noise level $\gamma=10^{-2}$ at red areas.}
	\label{fig:errortriangles}
\end{figure}

Finally, we also consider varying $\bfw$ when $\bfM$ is kept fixed at its true value. We generate $\bfd$ by using \eref{eq:fwd} and adding Gaussian noise with zero mean and standard deviation $\gamma=10^{-1}$ or $\gamma=10^{-2}$. For both noise levels, the reconstruction errors $e(\bfM,\bfw,\hat{\bfM},\hat{\bfw})$, based on the block coordinate descent MAP estimation method ($n_T=200$, $\eps_w = 10^{-3}$ and $n_I=10$), are shown side by side in Figure~\ref{fig:errortriangles}. As expected, larger reconstruction errors can be seen for the noise level $\gamma=10^{-1}$ in comparison to the case where the noise level is $\gamma=10^{-2}$. It is worth noting two instances where large reconstruction errors are present for $\gamma=10^{-2}$, namely when one component of $\bfw$ can be expressed as the sum of two other components, as represented by the vertical bright area at $\bfw_1$ = $0.5$, and when two components are equal, as shown by the bright areas near the edges of the depicted triangle. The large reconstruction errors reflect the problem of bi-dependency as described in Section~\ref{sub:map_estimation}.

\section{Discussion and conclusion}
\label{sec:summ}

In this paper, we present a mathematical formulation and computational framework for identifying strains of target microorganisms using PCR measurements from mixed samples. Extracting information about strains from mixed samples has the potential to reduce bias, time-to-results, and laboratory costs, and thus is critical for efficient screening.
Our method alleviates the need for culturing and isolating pathogens to produce detailed genetic information, which makes it attractive for public health applications involving samples composed of multiple strains of the same microorganism.
Epidemiological surveillance relies on the identification of microorganisms in samples, however, distinguishing multiple strains in mixed samples currently requires linking of locations~\cite{vollmers} or a prior dictionary of known strains~\cite{Zhu2018, Galinsky2015}. 
Our methods do not require these limiting assumptions and are thus more broadly applicable.

Our main contribution is the mathematical formulation of strain identification as an inverse problem that estimates a binary matrix encoding the strains and a vector modeling their relative contributions to the measured data. The resulting problem is highly underdetermined and, also due to the presence of the binary constraints, challenging to solve. We propose several efficient methods inspired by structurally similar problems such as blind source separation~\cite{talwar1994,talwar1996,diamantaras2000,behr2017}, non-negative matrix factorization~\cite{slawski2013} and blind deconvolution~\cite{YouKaveh1996, ChanWong1998, ChanWong2000} but also leveraging result from mixed-integer programming. 

Following a Bayesian approach, we derive a posterior density where prior information is incorporated to limit the underdetermined nature of the problem. The prior on the frequency vector enforces the non-negativity and sum-to-one properties, as well as a decreasing order to limit ambiguity. The prior on the strain matrix represents the binary constraints.

We propose efficient computational methods for exploring the posterior distribution. First, using block coordinate descent, we approximately solve the nonlinear mixed-integer problem arising in MAP estimation from different starting guesses to identify local and global modes. We exploit the fact that the optimization problem for the binary matrix decouples across rows to obtain a scheme whose complexity is linear with respect to the number of measurements and exponential in the number of strains to be recovered. Since the latter is relatively small in the target application, we can use full enumeration in this step. 
Second, we derive a convex re-formulation of the problem. This approach is less scalable but provides a lower bound for the negative log-likelihood that can be used to certify the optimality.
Third, we propose an efficient numerical integration technique for estimating the conditional mean and standard deviation of the posterior. 

As shown in our numerical examples on synthetic and experimental data with available ground truths, these methods allow one to discover the ambiguity of the problem at hand and capture uncertainty in the solution. Developing more scalable and accurate techniques to quantify the uncertainty by sampling from the multimodal posterior is a subject of future work. 

Our work paves the way for fast and inexpensive species-specific differentiation of strains of targeted microorganisms through DNA barcoding and whole genome multilocus sequence typing,
enabling epidemiologists and public health officials to conduct more granular tracking of pathogens and surveillance of infectious diseases.

\section{Acknowledgements}
\label{sec:acknowledgements}
We thank Ya Ping Shi at Malaria Branch, Division of Parasitic Diseases and Malaria, Center for Global Health, Centers for Disease Control and Prevention (CDC), U.S.A. for introducing us the need of new approach for molecular strain identification and providing initial funding.
The authors would also like to thank Felix Lucka (CWI, Netherlands and UCL, UK) for fruitful discussions and his useful suggestions.
This work is partially supported by Advanced Molecular Detection fund from the CDC and the National Science Foundation (NSF), awards DMS\#1522599 and CNS\#1553579.

\section{Bibliography}
\label{sec:bibliography}
\bibliography{MultiStrainInfection}

\begin{thebibliography}{10}

\bibitem{Galinsky2015}
K~Galinsky, C~Valim, A~Salmier, B~de~Thoisy, L~Musset, E~Legrand, et~al.
\newblock {COIL}:\ a methodology for evaluating malarial complexity of
  infection using likelihood from single nucleotide polymorphism data.
\newblock {\em Malaria Journal}, 14(1):4, 2015.

\bibitem{tenover1997}
FC~Tenover, RD~Arbeit, and RV~Goering.
\newblock How to select and interpret molecular strain typing methods for
  epidemiological studies of bacterial infections: a review for healthcare
  epidemiologists.
\newblock {\em Infection Control \& Hospital Epidemiology}, 18(6):426--439,
  1997.

\bibitem{Zhu2018}
SJ~Zhu, J~Almagro-Garcia, and G~McVean.
\newblock Deconvolution of multiple infections in \emph{{P}lasmodium
  falciparum} from high throughput sequencing data.
\newblock {\em Bioinformatics}, 34(1):9--15, 2018.

\bibitem{daniels2008}
R~Daniels, SK~Volkman, DA~Milner, N~Mahesh, DE~Neafsey, DJ~Park, et~al.
\newblock A general {SNP}-based molecular barcode for {P}lasmodium falciparum
  identification and tracking.
\newblock {\em Malaria Journal}, 7(1):223, 2008.

\bibitem{eurosurveillance}
C~Nadon, I~Van~Walle, P~Gerner-Smidt, J~Campos, I~Chinen, J~Concepcion-Acevedo,
  et~al.
\newblock {PulseNet International: Vision for the implementation of whole
  genome sequencing (WGS) for global food-borne disease surveillance}.
\newblock {\em Eurosurveillance}, 22(23), 2017.

\bibitem{maiden2013}
MC~Maiden, MJ~Jansen~van Rensburg, JE~Bray, SG~Earle, SA~Ford, KA~Jolley, and
  ND~McCarthy.
\newblock {MLST} revisited: the gene-by-gene approach to bacterial genomics.
\newblock {\em Nature Reviews Microbiology}, 11(10):728--736, 2013.

\bibitem{quainoo2017}
S~Quainoo, JPM Coolen, SAFT van Hijum, MA~Huynen, WJG Melchers, W~van Schaik,
  and HFL Wertheim.
\newblock Whole-genome sequencing of bacterial pathogens: the future of
  nosocomial outbreak analysis.
\newblock {\em Clinical Microbiology Reviews}, 30(4):1015--1063, 2017.

\bibitem{randall}
RS~Singer, AE~Mayer, TE~Hanson, and RE~Isaacson.
\newblock Do microbial interactions and cultivation media decrease the accuracy
  of {S}almonella surveillance systems and outbreak investigations?
\newblock {\em Journal of Food Protection}, 72(4):707--713, 2009.

\bibitem{dopfer2008assessing}
D~D{\"o}pfer, W~Buist, Y~Soyer, MA~Munoz, RN~Zadoks, L~Geue, and B~Engel.
\newblock Assessing genetic heterogeneity within bacterial species isolated
  from gastrointestinal and environmental samples: how many isolates does it
  take?
\newblock {\em Applied and Environmental Microbiology}, 74(11):3490--3496,
  2008.

\bibitem{JacobEtAl2011}
ME~Jacob, KM~Almes, X~Shi, JM~Sargeant, and TG~Nagaraja.
\newblock Escherichia coli {O157: H7} genetic diversity in bovine fecal
  samples.
\newblock {\em Journal of Food Protection}, 74(7):1186--1188, 2011.

\bibitem{SacchiEtAl2011}
CT~Sacchi, LO~Fukasawa, MG~Gon{\c c}alves, MM~Salgado, KA~Shutt,
  TR~Carvalhanas, et~al.
\newblock Incorporation of real-time {PCR} into routine public health
  surveillance of culture negative bacterial meningitis in {S{\~a}o Paulo,
  Brazil}.
\newblock {\em PLoS ONE}, 6(6):e20675, 2011.

\bibitem{LangleyEtAl2015}
G~Langley, J~Besser, M~Iwamoto, et~al.
\newblock Effect of culture-independent diagnostic tests on future emerging
  infections program surveillance.
\newblock {\em Emerging Infectious Diseases}, 21(9):1582--1588, 2015.

\bibitem{vollmers}
J~Vollmers, S~Wiegand, and A-K Kaster.
\newblock Comparing and evaluating metagenome assembly tools from a
  microbiologist's perspective - not only size matters!
\newblock {\em PLoS ONE}, 12(1):e0169662, 2017.

\bibitem{KaipioSomersalo2006}
J~Kaipio and E~Somersalo.
\newblock {\em {Statistical and Computational Inverse Problems}}, volume 160 of
  {\em Applied Mathematical Sciences}.
\newblock Springer Science {\&} Business Media, New York, 2006.

\bibitem{CalvettiSomersalo2007}
D~Calvetti and E~Somersalo.
\newblock {\em {An Introduction to Bayesian Scientific Computing}}, volume~2 of
  {\em Ten Lectures on Subjective Computing}.
\newblock Springer Science {\&} Business Media, New York, NY, November 2007.

\bibitem{diamantaras2000}
KI~Diamantaras and E~Chassioti.
\newblock Blind separation of {$N$} binary sources from one observation: {A}
  deterministic approach.
\newblock In {\em International Workshop on Independent Component Analysis and
  Blind Signal Separation}, pages 93--98, 2000.

\bibitem{diamantaras2006cluster}
KI~Diamantaras.
\newblock A clustering approach for the blind separation of multiple finite
  alphabet sequences from a single linear mixture.
\newblock {\em Signal Processing}, 86(4):877--891, 2006.

\bibitem{belouchrani1997}
A~Belouchrani, K~Abed-Meraim, J-F Cardoso, and E~Moulines.
\newblock A blind source separation technique using second-order statistics.
\newblock {\em IEEE Transactions on Signal Processing}, 45(2):434--444, 1997.

\bibitem{talwar1994}
S~Talwar, M~Viberg, and A~Paulraj.
\newblock Blind estimation of multiple co-channel digital signals using an
  antenna array.
\newblock {\em IEEE Signal Processing Letters}, 1(2):29--31, 1994.

\bibitem{behr2017}
M~Behr and A~Munk.
\newblock Identifiability for blind source separation of multiple finite
  alphabet linear mixtures.
\newblock {\em IEEE Transactions on Information Theory}, 63(9):5506--5517,
  2017.

\bibitem{lin2007}
C-J Lin.
\newblock Projected gradient methods for nonnegative matrix factorization.
\newblock {\em Neural Computation}, 19(10):2756--2779, 2007.

\bibitem{slawski2013}
M~Slawski, M~Hein, and P~Lutsik.
\newblock Matrix factorization with binary components.
\newblock In {\em Advances in Neural Information Processing Systems}, pages
  3210--3218, 2013.

\bibitem{YouKaveh1996}
YL~You and M~Kaveh.
\newblock {A regularization approach to joint blur identification and image
  restoration}.
\newblock {\em IEEE Transactions on Image Processing}, 5(4):416--428, 1996.

\bibitem{ChanWong1998}
TF~Chan and CK~Wong.
\newblock {Total variation blind deconvolution}.
\newblock {\em IEEE Transactions on Image Processing}, 7(3):370--375, 1998.

\bibitem{ChanWong2000}
TF~Chan and CK~Wong.
\newblock {Convergence of the alternating minimization algorithm for blind
  deconvolution}.
\newblock {\em Linear Algebra and its Applications}, 316(1-3):259--285, 2000.

\bibitem{LairdEtAl2006}
CD~Laird, LT~Biegler, and B~van Bloemen~Waanders.
\newblock {Mixed-integer approach for obtaining unique solutions in source
  inversion of water networks}.
\newblock {\em Journal of Water Resources Planning and Management},
  132(4):242--251, 2006.

\bibitem{BelottiEtAl2013}
P~Belotti, C~Kirches, S~Leyffer, J~Linderoth, J~Luedtke, and A~Mahajan.
\newblock {Mixed-integer nonlinear optimization}.
\newblock {\em Acta Numerica}, 22:1--131, May 2013.

\bibitem{talwar1996}
S~Talwar, M~Viberg, and A~Paulraj.
\newblock Blind separation of synchronous co-channel digital signals using an
  antenna array. {I}. {A}lgorithms.
\newblock {\em IEEE Transactions on Signal Processing}, 44(5):1184--1197, 1996.

\bibitem{leyffer2012minotaur}
A~Mahajan, S~Leyffer, JT~Linderoth, JR~Luedtke, and T~Munson.
\newblock Minotaur: A mixed-integer nonlinear optimization toolkit.
\newblock 2017.

\bibitem{gurobi}
{Gurobi Optimization, Inc.}
\newblock Gurobi optimizer reference manual, 2016.

\bibitem{li2003}
Y~Li, A~Cichocki, and L~Zhang.
\newblock Blind separation and extraction of binary sources.
\newblock {\em IEICE Transactions on Fundamentals of Electronics,
  Communications and Computer Sciences}, 86(3):580--589, 2003.

\bibitem{NocedalWright2006}
J~Nocedal and S~Wright.
\newblock {\em {Numerical Optimization}}.
\newblock Springer Series in Operations Research and Financial Engineering.
  Springer Science {\&} Business Media, New York, 2006.

\bibitem{Hamarik2016}
U~H{\"a}marik, B~Kaltenbacher, U~Kangro, and E~Resmerita.
\newblock Regularization by discretization in {B}anach spaces.
\newblock {\em Inverse Problems}, 32(3):035004, 2016.

\bibitem{Hansen2010}
PC~Hansen.
\newblock {\em {Discrete inverse problems}}, volume~7 of {\em Fundamentals of
  Algorithms}.
\newblock Society for Industrial and Applied Mathematics (SIAM), Philadelphia,
  PA, 2010.

\bibitem{cplex}
{IBM Corp.}
\newblock {IBM ILOG CPLEX Optimization Studio: CPLEX User's Manual}, 2016.

\bibitem{Shabbir2013}
A~Gupte, S~Ahmed, MS~Cheon, and S~Dey.
\newblock Solving mixed integer bilinear problems using {MILP} formulations.
\newblock {\em SIAM Journal on Optimization}, 23(2):721--744, 2013.

\bibitem{Gupte2017}
A~Gupte, S~Ahmed, SS~Dey, and MS~Cheon.
\newblock Relaxations and discretizations for the pooling problem.
\newblock {\em Journal of Global Optimization}, 67(3):631--669, 2017.

\bibitem{Schumacher2017}
KM~Schumacher, RL-Y Chen, and AEM Cohn.
\newblock Transmission expansion with smart switching under demand uncertainty
  and line failures.
\newblock {\em Energy Systems}, 8(3):549--580, 2017.

\bibitem{McCormick1976}
GP~McCormick.
\newblock Computability of global solutions to factorable nonconvex programs:
  Part {I} --- convex underestimating problems.
\newblock {\em Mathematical Programming}, 10(1):147--175, 1976.

\bibitem{JacodProtter2004}
J~Jacod and P~Protter.
\newblock {\em Probability Essentials}.
\newblock Springer Science \& Business Media, 2004.

\bibitem{BezansonEtAl2012}
J~Bezanson, S~Karpinski, VB~Shah, and A~Edelman.
\newblock Julia: A fast dynamic language for technical computing.
\newblock {\em arXiv:1209.5145}, 2012.

\bibitem{Dunning2015}
I~Dunning, J~Huchette, and M~Lubin.
\newblock {JuMP}: A modeling language for mathematical optimization.
\newblock {\em SIAM Review}, 59(2):295--320, 2017.

\end{thebibliography}
\bibliographystyle{unsrt}

\end{document}